\documentclass[12pt]{article}
\usepackage{latexsym}
\usepackage{amsfonts}
\usepackage{acronym}
\newcommand{\be}{\begin{equation}}
\newcommand{\ee}{\end{equation}}
\newcommand{\bea}{\begin{eqnarray}}
\newcommand{\eea}{\end{eqnarray}}
\makeatletter
\@addtoreset{equation}{section}
\def\theequation{\thesection.\arabic{equation}}
 \makeatother
\begin{document}
\title{Monopole Vacuum in Non-Abelian Theories}
\author
{{\bf L.~D.~Lantsman}\\
 Wissenschaftliche Gesellschaft, bei
 J$\ddot u$dische Gemeinde  zu Rostock,\\Wilhelm-K$\ddot u$lz Platz,6;
 \\18055, Rostock, Germany; \\ llantsman@freenet.de \\\\
{ $~~~~$\bf V.~N.~Pervushin} \\
  Bogoliubov Laboratory of Theoretical Physics, \\  Joint Institute for
Nuclear
Research, \\ 141980 Dubna, Russia;\\ pervush@thsun1.jinr.ru}
\maketitle
\medskip
\begin{abstract} It is shown that, in the theory of interacting Yang -Mills fields and a Higgs field, there is a
topological degeneracy of Bogomol'nyi-Prasad-Sommerfield (BPS)
monopoles and that there arises, in this case, a chromoelectric
monopole characterized by a new topological variable that
describes transitions between topological states of the monopole
in the Minkowski space (in just the same way as an instanton
describes such transitions in the Euclidean space). The limit of
an infinitely large mass of the Higgs field at a finite density of
the BPS monopole is considered as a model of the stable vacuum in
the pure Yang-Mills theory.  It is shown that, in QCD, such a
monopole vacuum may lead to a rising  potential, a topological
confinement and an additional mass of the $\eta_0$  meson. The
relationship between the result obtained here for the  generating
functional of perturbation theory and  Faddeev-Popov integral is
discussed.
\end{abstract}
\newpage
\section{ Introduction and formulation of  problem.}
The problem of choosing, in a non-Abelian theory,
a physical vacuum and variables that adequately reflect
 the topological properties of the manifold of initial
data for non-Abelian fields \cite{Bel,Jack,Hooft1} in the Minkowski space is still considered as one of the most important problems in these realms. There are reasons to believe
that solving this problem will contribute to obtaining,
within QCD, a deeper insight into the nature of the
confinement, hadronization and  spontaneous
break-down of scale invariance in the infrared region. \par
The present study is devoted to employing  monopole solutions \cite{BPS} to the equations of a
non-Abelian
theory for the purpose to construct a model of a topologically
invariant vacuum of  the Yang-Mills (YM) theory in the Minkowski
space. The respective Lagrangian density of that  theory has
the form
\be  \label{L}
{\cal L}= -\frac {1}{4} G_{\mu \nu}^a G_a^{\mu \nu },\ee
where
\be  \label{L1}
G_{\mu \nu}^a= \partial _\mu
A_\nu^a -\partial _\nu A_\mu ^a+ g \epsilon ^{abc} A_{b\mu} A_{c\nu}.
\ee
>From the mathematical point of view, the main
problem in a theory of gauge fields is to find general
solutions to the equations of that theory,
\be  \label{D}
D _\mu ^{ab} G_b^{\mu \nu}=0\quad (D^{ab}_ \mu= \delta ^{ab}\partial_\mu+ g
\epsilon ^{acb}A_{c\mu}),
\ee
and to construct, in order to describe processes in
terms of probability amplitudes normalized per time
and spatial-volume units, the generating functional for
the appropriate $S$-matrix
in the class of functions such that the
energy density is finite \cite{NN1,A.I.,Schweber}. In QED such functions have  the  $O(1/r^{1+m})$ behaviour at the spatial infinity.
They
are referred to as monopoles if $m=0$ and as multipoles if $m> 0$. \par
 Solving differential equations in theoretical
physics presumes specifying initial data. Such initial data
are measured by a set of physical instruments with
which one associates a reference frame. In the present
study we will consider reference frames that are  determined by the timelike unit vector $l_\mu ^{(0)}=(1,0,0,0) $ and various Lorentz transformations of it, $l_\mu ^{(1)}$.\par
There are two types of the groups of transformations of the differential equations of a gauge theory.
There are relativistic transformations, which change
initial data (i.e. the appropriate reference frame) and gauge
transformations,
\be  \label{gauge}
{\hat A}^u_\mu(t; {\bf x}):=u(t; {\bf x})({\hat A}_\mu+\partial_\mu)u^{-1}(t,{\bf x}
); \quad {\hat A}_\mu= g \frac {\tau ^a}{2i}A_{a\mu};
\ee
which are associated with the gauge of physical fields
and which do not affect the readings of an instrument. \par
The set of equations (\ref{D})
is referred  to as a relativistic
covariant set of equations if the total manifold of its
solutions for each specific reference frame coincides
with its counterpart for any other reference frame (or isomorphic it) \cite{Schweber, Schwinger1,Zumino,Schwinger2,Polubarinov,Pervush1}.
\par
In each reference frame the set of all the equations is split into the equations of motion:  $D _\mu ^{ab} G_b^{\mu i}=0 ~(i=1,2,3)$; to solve these equations, it is necessary
to measure initial data, and the constraint equations $D _\mu ^{ab} G_b^{\mu 0}=0$, which relate initial data for the spatial
components of the fields involved to initial data for
their temporal components.
\footnote{
The both sets may intersect, as far as
the constraint equations are imposed onto the generalized velocities, i.e. onto the
first time derivatives of gauge fields (see, e.g., the theory
(1.11)- (1.12) in \cite{Gitman}). An important example of such intersection
is the equation on the temporal component of a YM field. It is the
motion equation and constraint simultaneously. We shall utilize this fact in the present
statement. \par
In the "pure" YM theory (see \cite{Gitman}, \S 16) temporal components
of  YM fields occupy a particular position, since they have no nonzero canonical
momenta:
$$E_0=\partial {\cal L}/\partial (\partial_0
 A_0)=0,$$
that contradict the commutation relations and Heisenberg's uncertainty principle  \cite{Pervush1}. \par
Such situation takes place in the gauge theories involving \cite{Gitman} degenerated Hessian matrices
$$ M_{ab}= \frac {\partial ^2 {\cal L}}{\partial \dot q ^a\partial\dot q ^b},$$
with $q ^i$ being   appropriate  degrees of
freedom in the gauge theory considered and $\dot q^i$ being their time derivatives. Maxwell electrodynamics  and non-Abelian models are patterns of that theories.} \par
In view of the said, Dirac \cite{Dir} and, after him, other authors of
the first classic studies devoted to quantizing   gauge
theories (see \cite{Heisenberg,Fermi}) removed temporal field components by  gauge transformations. In our case,  such a transformation,
\be  \label{hata}
\hat A^D_k = v({\bf x})T \exp \{\int  \limits_{t_0}^t d {\bar t}\hat A _0(\bar t,
{\bf x})\}({\hat A}_k+\partial_k) [v({\bf x})T \exp \{\int  \limits_{t_0}^t d {\bar t}
\hat A _0(\bar t,{\bf x})\}]^{-1}
\ee
(here the symbol $T$ denotes the time ordering of the
matrices under the exponential sign), specifies a non-Abelian analogue  of Dirac's variables:  apart from   arbitrary stationary matrices $ v({\bf x})$,  which are
considered as initial data, at  the time instant $t_0$,  for  solving the
equation $$U (A
_0+\partial) U^{-1}=0.$$
 At the level of Dirac's
variables,   Lorentz transformations of original fields become
non-linear, while the group of gauge transformations reduces to a
group of stationary transformations that specify the degeneracy of
initial data for physical  fields (including the classical vacuum
$A_0=A_i=0$, which is defined as the zero-energy state). By the
gauge fixation one means, in this case, the presetting initial
data in a perturbation theory as the transversality condition
\cite{Zumino}-\cite{Pervush1}. \par In the YM  theory the set of
stationary gauge transformations is the set of three-dimensional
paths in the space of the $SU_c(2)$ group that are broken down
into topological manifolds  characterized by integers (\it
Pontryagin degrees of a map\rm): \begin{eqnarray}    {\cal
N}[n]&=&- \frac {1}{24\pi ^2}\int d^3 x \epsilon ^{ijk} Tr
[v^{(n)}( {\bf x})\partial_i v^{(n)}({\bf x})^{-1} v^{(n)}({\bf
x})\partial_j v^{(n)}({\bf x})^{-1} v^{(n)}({\bf x})\partial_k
v^{(n)}({\bf x} )^{-1}]\nonumber\\\label{degree} &=&n \in{\bf Z}.
\end{eqnarray} Any degree of a map indicates how many times the
three-dimensional path $v( \mathbf{x})$ goes about the $SU_c(2)$
group manifold as the coordinate $\bf x_i$ runs over the entire
three-dimensional space where this coordinate is specified. \par
The condition (\ref{degree}) means that the complete set of
three-dimensional paths has the homotopic  group $\pi_3 (SU_c(2))
={\bf Z}$ and that all the   fields $v^{(n)}\partial_i v^{(n)-1}$
are given
 in the class of functions for which the integral  (\ref{degree}) takes finite (or countable) values.
This is the class of monopole functions \cite{Bel,Jack}. Naturally, the fields $A_i^D
(t,{\bf x})$ themselves also must
belong to this class of monopole functions
and have the $O(1/r^{1+m})$, $m> 0$, asymptotic behaviour. \par
Thus
our objective is to quantize non-Abelian
fields in the class of monopole functions that involve
a topological degeneracy. Such a quantization presumes the choice of Dirac's variables in which this
degeneracy occurs. \par
The  primary Hamiltonian quantization of non-Abelian gauge theories in terms of Dirac's
variables without allowing for their topological degeneracy was due to Schwinger \cite {Schwinger2},
who proved the
relativistic covariance of the transverse gauge at the
level of commutation relations for the generators of
the algebra of the Poincare group that were constructed in the above class of functions. This Hamiltonian quantization of non-Abelian fields was then reproduced by Faddeev \cite {Fadd1}, who employed the method of a
path integral $Z_{l^{(0)}}$ explicitly dependent on a reference
frame.  There
was shown in \cite {Nguen1} that the  relativistic transformation at the level of fundamental operator
quantization by  Schwinger \cite { Schwinger2} corresponds to
the relativistic transformation $l^{(0)}\to l^{(1)}$
of the time
axis on which depends the path integral $Z_{l^{(0)}}$.
 The
dependence of this integral on a reference frame is
called an implicit relativistic covariance.
\footnote{The choice of the time axis for such an integral
was discussed in the works  \cite{Ilieva1,Yura1,Pervush2}.}\par
That the path integral $Z_{l^{(0)}}$
is independent on a reference frame for on-shell amplitudes of elementary-particle scattering was first discovered by Feynman \cite {Feynman1} and was proven by Faddeev \cite{Fadd1}
as a
validation of the heuristic Faddeev-Popov  (FP) path integral \cite{FP1}.
This integral was proposed as a generating functional of unitary perturbation theory for any gauges, including those that are independent of a
reference frame. Schwinger noticed that gauges that
are independent of a reference frame may be physically
inadequate to the fundamental operator quantization;
i.e. they may distort the spectrum of the original
system.
\footnote{"We reject all Lorentz gauge formulations as unsuited to the role of providing
the fundamental operator quantization"\cite{Schwinger2}.}\par
In the present study we verify this
Schwinger's statement in a non-Abelian theory, answering the question
concerning the spectrum of a theory quantized in the
class of monopole functions that involves a topological degeneracy of initial data and the question
concerning the relationship between the fundamental quantization and the heuristic FP integral in
a gauge that is independent on a reference frame. \par
The ensuing exposition is organized as follows.
Section 2 is devoted to  describing the topological degeneracy of known monopole solutions and to considering zero modes of the constraint equation. In
Section 3 we examine the limiting transition to the "pure"
YM theory having a monopole vacuum. In
Section 4 we analyse the $U(1)$-problem. In the Conclusion we discuss the connection with the
FP integral.
\section{Topological degeneracy of BPS monopole.}
Let us consider the well-known example of interacting YM and scalar Higgs fields for the case where there is a spontaneous breakdown
of the initial $SU(2)$ gauge symmetry. This situation is described by the Lagrangian density \cite{BPS}
\be
{\cal L} = -\frac {1}{4} G_{\mu \nu}^a G^{\mu \nu}_a +
\frac {1}{2} (D_\mu \phi^a)(D^\mu \phi_a)-
\frac {\lambda}{4} (\frac {m^2}{\lambda}-\phi ^2)^2,
\label {Lag}
\ee
where
\be
G_{\mu \nu}^a = \partial _\mu A_\nu^a- \partial _\nu A_\mu^a +
g \epsilon^{abc}A_{b\mu}A_{c\nu}
\label {tens}
\ee
is the strength tensor of our non-Abelian field and
$\phi^a$  $(a=1,2,3)$ is a scalar field forming a triplet of
the adjoin representation of the $SU(2)$ gauge group. The
potential energy in the YM model considered depends on the square of the isotopic vector $\phi^a$,
while the covariant derivative of a  YM field has the
form
\be
D^\mu \phi_a =\partial^\mu \phi_a +g \epsilon_{abc} A^{b\mu}\phi^c.
\label{der}
\ee
The Lagrangian density ${\cal L}$ possesses a manifest
gauge invariance under transformations of the $SU(2)$ group.\par
The classical vacuum is defined as the asymptotic
solution
\be
 \label {asol}
r = \vert {\bf x} \vert \to \infty ,\quad E \to {\rm min} ~E,
\ee
providing the minimum of the field energy $E$.\par
For $m^2\geq 0$ and $\lambda \geq 0 $, the Minkowskian YM vacuum loses its initial $SU(2)$ gauge symmetry of the Lagrangian (\ref {Lag}), i.e.
\be
 \label {sym}
r = \vert {\bf x} \vert \to \infty ,\quad   \phi_a \to n_a \frac {m}{\sqrt {\lambda }},
\ee
 with  $n_a $ being an arbitrary unit vector ($ \vert {\bf n} \vert =1$) in the isotopic space. \par
One may be shown (see \S 10.4 in \cite{Ryder}) that the vacuum
manifold $M$, i.e. the surface where  the  Higgs potential
$$V\equiv \frac {\lambda}{4} (\frac {m^2}{\lambda}-\phi ^2)^2$$
reaches its (topologically degenerated) minimum in the isotopic space of Higgs vectors $\phi^a$, is invariant under the residual gauge symmetry group $U(1) \equiv H$. \par
In this case the vacuum
manifold $M$ possesses the manifest geometry of the two-sphere $S^2$:
\be
M= S^2=\{\phi= a;\quad a ^2=m^2/ \lambda\};
\label {sphere}
\ee
of the radius $a$
\footnote{
Let us denote $SU(2)\equiv G$.
Since the vacuum manifold $M$ is invariant with respect to  the
transformations of $H$, it can be subdivided onto the
$H$-orbits of its points: $M=\sum \limits_i H M_i$.
As far as $H\subset G$, the vacuum manifold $M$ is
given as a \it set of those transformations of  $G$ which
do not  belong to $H$\rm. Thus $$M=G/H= \sum \limits_i H M_i$$
and
$$G=H+ G/H.$$
 Such sketch is correct for each spontaneous breakdown of initial symmetry in gauge theories.}. \par
If we  consider the maps of the sphere $M$,  in the isotopic space of Higgs vectors, into the spatial sphere
$S^2:=\{\vert{\bf n} \vert=1\}$ as
 $ {\bf r} \to \infty$,  we obtain the chain of  topological
equalities:
\be
\label{top2}
\pi_2 S^2= \pi_3 (SU(2))=\pi_1(U(1))=\pi_1S^1={\bf Z}.
\ee
Just this nontrivial topology determines magnetic charges associated with the residual $U(1)$ gauge symmetry in the Minkowskian YM  theory (they themselves  point to
an electromagnetic theory).\par
In the Standard Model a non-Abelian vector field
develops a mass in precisely this way. Usually, a quantum-field theory is then constructed as a perturbation
theory over this vacuum  in the class of function with finite energy densities.\par
In addition to the Minkowskian YM vacuum $M$, (\ref{sphere}),
there are
monopole solutions in the system described by the
Lagrangian density  (\ref{Lag}). These are solutions having
the $O(1/r)$ type  behaviour at the spatial infinity:
\be
\label{dis}
r \to \infty; \quad   \phi_a -  n_a \frac {m}{\sqrt {\lambda }}= O(1/r), \quad
A_i^a =  O(1/r).
\ee
A pattern of such monopole solutions is the Bogomol'nyi-Prasad-Sommerfield (BPS)
solution \cite{BPS}, in the zero topological sector ($n=0$), for the Minkowskian (YM-Higgs) vacuum:
\be
\label{sc monopol}
\phi^a =
 \frac{ x^a}{gr} f_0^{BPS}(r)~,~~~~~~~~~~~~
 f_0^{BPS}(r)=\left[  \frac{1}{\epsilon\tanh(r/\epsilon)}-\frac{1}{r}\right],
\ee \be
\label{YM monopol}
A^a_i(t,{\bf x})\equiv\Phi^{aBPS}_i({\bf x}) =\epsilon_{iak}\frac{x^k}{gr^2}f^{BPS}_{1}(r),~~~~~~~~
  f^{BPS}_{1}= \left[1 -
 \frac{r}{\epsilon \sinh(r/\epsilon)}\right],
\ee
which was obtained in the Bogomol'nyi-Prasad-Sommerfield (BPS)
limit
\be
\label{lim}
\lambda\to 0,~~~~~~m\to 0 ;~~~~~~~~~~
~~~~~\frac{1}{\epsilon}\equiv\frac{gm}{\sqrt{\lambda}}\not =0,
\ee
and is compatible with the topology (\ref {top2})
in  a finite spatial volume in the Minkowski space
\footnote{ The statement that the solutions (\ref {sc monopol}), (\ref{YM monopol}) are regular in the finite
volume implies that we should consider
the topology (\ref{top2}) and  vacuum manifold $M$,
(\ref{sphere}), taking account of this finite spatial
volume. When we wish to adapt our theory to needs of
QCD (in Section 3 we shall see how to do this), the spatial
volume specified by the typical hadronic
size, $\sim 1$ fm. ($\sim 5$ GeV$^{-1}$), is quite sufficient for
our purposes.
}.\par
The BPS vacuum solution  (\ref {sc monopol}), (\ref{YM monopol})
is a monopole solution satisfying the YM
equations and associated with the (topologically degenerated) lowest bound of the "(YM-Higgs)" energy \cite{Al.S.} (often referred to  as the \it Bogomol'nyi bound\rm: see, e.g. \cite{Hooft1}).
\be
\label{Emin}
E_{min}= 4\pi {\bf m} \frac {a}{g},~~~~~~~~~~~~\,\,
~~~~~a=\frac{m}{\sqrt{\lambda}}:
\ee
with ${\bf m}$ being   magnetic charges associated with vacuum BPS monopoles. \par
In the limit specified in (\ref {lim}), the BPS solution  (\ref {sc monopol}), (\ref{YM monopol})  involves the minimum (vacuum) "magnetic" energy that is proportional to
\be
 \label{mine}
\int  d^3x [B_i ^a B^i_a] =  4\pi \frac{gm}{g^2\sqrt{\lambda}}=
\frac{4\pi}{g^2 \epsilon},
\ee
where $B_i ^a$ is the vacuum "magnetic" tension,
\be
 \label{mf}
B_i^a  (\Phi_k^{cBPS}) =  \epsilon_{ijk}(\partial_j \Phi_k^{aBPS}+ \frac{g}{2}
\epsilon^{abc} \Phi_j^{bBPS}\Phi_k^{cBPS}). \ee
The both formulas: (\ref {Emin}), for the Bogomol'nyi bound $ E_{min}$ of the "(YM-Higgs)" energy, and (\ref{mf}), for the vacuum "magnetic" energy, are closely related  each  with other \cite{Al.S.}. \par
That the vacuum "magnetic" energy (\ref{mf}) corresponds to the Bogomol'nyi bound $ E_{min}$ of the "(YM-Higgs)" energy
is
ensured by the requirement that the quested vacuum "magnetic" field $\bf B$ would be
of a potential character:
\be
 \label{Bog}
B_i^a  (\Phi_k^{cBPS}) =  D _i^{ab} (\Phi_k^{cBPS}) \phi _b,
\ee
where the covariant derivative is specified by the equation (\ref{der}).
Just  this condition (which ensures, as was
indicated immediately above, the potential character
of the vacuum "magnetic" field $\bf B$): it is referred to as the
Bogomol'nyi equation \cite{BPS,Al.S.,Gold},  will play an important role in our construction of the stable vacuum of a non-Abelian theory (see below). \par
We will  show now that the equation of potentiality (\ref{Bog})
means a topological degeneracy of fields under the
stationary gauge transformations
\be
\label{degeneration}
{\hat A}^{(n)}_i (t_0,{\bf x}) = v^{(n)}({\bf x}) [{\hat A}_i^{ (0)}(t_0,{\bf x}) +
\partial _i] v^{(n)}({\bf x})^{-1}.
\ee
Dynamical fields can be represented in the form of
the sum of the vacuum BPS monopole $\Phi_i ^{(0) BPS }({\bf x})$ and perturbations $\bar A _i^{(0)}$:
\be
\label{s}
\hat A_i^{ (0)}(t,{\bf x}) =\Phi_i^{(0) BPS}({\bf x}) +
{\hat{\bar A}}_i^{ (0)} (t,{\bf x}).
\ee
Perturbations are considered as weak multipoles
 \cite {Gold}:
\be
\label{asym}
{\bar  A}_i (t,{\bf x})\vert _{assymptotics}= O (\frac {1}{r^{l+1}})\quad (l>1).
\ee
In the lowest order of the perturbation theory the
equation for  the temporal component  of a YM field (it is \it the motion equation and constraint
simultaneously\rm),
\be
\label{Gauss}
[D^2(A)]^{ac}A^0_c =[ D^{ac}_i(A)\partial_0 A_{c}^i],
\ee
in Dirac's variables $A_0^{Dc}=0$ assumes the form
$$\partial _t A^{a\parallel}[A_i^{c(0)}(t,{\bf x})] =0,$$
\be
\label{init}
A^{a\parallel}[A_i^{c(0)}] \equiv  [ D^{ac}_i(\Phi^{(0)BPS}) A_i^{c(0)}],
\ee
and implies that the time derivative of the longitudinal
fields $A^{a\parallel}$ vanishes.  This equation can be solved if we
have initial data at our disposal.  We assume that there
are no longitudinal fields at the initial instant of time;
that is,
\be
\label{Aparallel}
A^{a\parallel}\equiv [D^{ac}_i(\Phi ^{(0)BPS})A_i^{c (0)}]=0\vert _{t=t_0}.
\ee
We refer to this condition as the covariant Coulomb
gauge. There arises the question of the degree of
arbitrariness in the specification of  Dirac's variables associated with this
gauge, since it should be recalled that they are defined
apart from stationary gauge transformations. \par
In order to answer this question, we make these
transformations:
\be
\label{degeneration1}
{\hat A}^{(n)}_i= v^{(n)} ({\hat A}_i^{ (0)}+
\partial _i)v^{(n)-1},\quad v^{(n)}({\bf x})=
\exp [n\hat \Phi _0({\bf x})],
\ee
and require that, upon the transformations (\ref{degeneration1}), the
fields
\be
\label{s1}
\hat A_i^{ (n)}(t,{\bf x}) =\Phi_i^{ (n)BPS}({\bf x}) +
{\hat{\bar A}}_i^{ (n)} (t,{\bf x})
\ee
satisfy the same covariant Coulomb gauge:
\be
\label{transv} D_i^{ab} (\Phi _k^{(n)BPS}){\bar A}^{i(n)}_b =0.
\ee
>From the last condition of the gauge conservation
we
then obtain the so-called Gribov (ambiguity) equation \cite{Gribov}
for the
phases of gauge transformations:
\be
\label{Gribov.eq}
[D^2 _i(\Phi _k^{BPS})]^{ab}\Phi_{0b} =0. \ee  The Gribov equation (\ref {Gribov.eq}) mathematically coincides with the Bogomol'nyi equation (\ref{Bog})
for the scalar field;
the latter implies the  potentiality   of the vacuum "magnetic" field $\bf B$ induced by vacuum BPS monopoles. Therefore, the Gribov equation (\ref {Gribov.eq})
has a nontrivial
solution in the form of a BPS monopole of the (\ref {sc monopol}) type:
\be
\label{phase}
{\hat \Phi}_0= -i\pi \frac {\tau ^a x_a}{r}f_{01}^{BPS}(r),
\quad f_{01}^{BPS}(r)=[\frac{1}{\tanh (r/\epsilon)}-\frac{\epsilon}{r} ].
\ee
Thus we have shown that vacuum YM BPS
monopoles  and transverse gauge physical fields belonging to the zero topological sector have their Gribov's copies in the form of  the topological degeneracy (\ref  {degeneration1}).\par
It should be recalled that a topological degeneracy
is associated primarily with a classical vacuum of
zero energy, where this degeneracy is characterized
by the Pontryagin index or by the Chern-Simons
functional \cite{Bel}
(which we consider in a finite space-time
of the volume $V$ within the time interval $t_{in}<t<t_{out}$):
\be
 \label{Ch-S}
\nu[A]=\frac {g^2}{16\pi ^2}\int\limits_ {t_{in}} ^{t_{out}} dt\int \limits_V d^3 x G_{\mu \nu}^a
{\tilde G}^{\mu \nu}_a = X[A^D_{out}]-X[A^D_{in}],
\ee
where (see (10.93) in \cite {Ryder})
\be
\label{wind}
X[A]=-\frac {1}{8\pi ^2}\int \limits_{V} d^3 x \epsilon ^{ijk}Tr [{\hat A}_i \partial_j{\hat A}_k-
\frac {2}{3}{\hat A}_i{\hat A}_j{\hat A}_k]
\ee
is a topological functional of gauge fields that is reduced to an integer
for a purely gauge field characterized by the degree of a map (\ref{degree}). \par
The functional (\ref{wind})
degenerates the quantum
wave function
\be
\label{wave}
\Psi _{ins}[A]= \exp \{\pm \frac{8\pi^2}{g^2} X[A]\}
\ee
as an exact solution to the Schr$\ddot o$dinger equation \cite{Jack,Pervush3}
\be
\label{Schrod}
{\hat H} \Psi _{ins}[A]=0 ;\quad  {\hat H}=\frac {1}{2}
\int d^3 [{\hat E}^2+{\hat B}^2] ;\quad {\hat E} = \frac {\delta}{i\delta A },
\ee
at the zero energy, $H=0$.
In just the same way as
the oscillator wave function at the zero energy,
$$({\hat p}^2+q^2)\Psi[q]=0,$$
this wave function is nonnormalizable,
which means that the corresponding eigenenergy $H$ belongs to unphysical values in the
energy-momentum spectrum. This fact may suggest that the instanton corresponding to transitions between vacua characterized by unphysical zero values of energy is itself
an unphysical solution. \par
Moreover, the formula (\ref  {wave})  for the instanton wave function
  implies that the  topological motion $X[A]$ is a functional of local degrees of freedom, that are denoted by $A$. In this case the operator
of local gauge transformations
$${\hat T}X[A]=X[A]+1$$
does
not commute with the Hamilton operator $\hat H$. \par
One of the
simplest ways to remove all of these flaws, including
the nonnormalizability of the wave function
\be
\label{plan}
\Psi _{ins}[A]=\exp \{iP_X X[A] \}
\ee
and the unphysical values of the energy and the momentum  $P_X$ of the topological motion,
\be
\label{spe}
H=0 , \quad  P_X = \pm \frac {8\pi ^2~i}{g^2}, \ee
consists in  separating the topological motion from the
local variables via the introduction of an independent topological
degree of freedom $N(t)$ by means of the
gauge transformation \cite{Pervush3,Pervush4}
\be
\label{d.v.} \hat A_i^{(N)}=\exp [N(t)\hat \Phi_0 ({\bf x})]
[\hat A^{(0)}_i+\partial _i]\exp [-N(t)\hat \Phi_0 ({\bf x})]~.
\ee
By means of a direct calculation, it can be proven \cite{David1} that, for the vacuum BPS monopoles $ \Phi_i ^{(n)}$,  this degree of freedom is
completely separated from the local degrees of freedom that are specified in the class of multipole
functions:
\be
\label{sep}
X[A_i^{(N)}]=   X[A_i^{(0)}]+  N(t).
 \ee
In this case the instanton wave function (\ref {plan}) acquires a new independent degree of freedom,
\be
\label{tri}
\Psi _{in}[A^N]=\exp\{iP_{ N}X[A^N] \}= \exp\{iP_{ N}(X[A^{(0)}] +N) \},
 \ee
and describes the topological motion of this degree
of freedom at physical values of the momentum $P_N$. An independent topological motion arises as an inevitable
consequence of a general solution to the
equation $D_\mu G^ {\mu0}= 0$ for  temporal YM components \cite{Pervush3,Pervush4}. This equation has the form
\be \label{time}
[D_k(A)]^{ab}[D^k(A)]_{bc}A_{0c} = D^{a}_{ci}(A)\partial_0 A^{ci},
\ee
with the initial data being those that correspond to the vacuum BPS monopole:
\be
\label{inid}
\partial _0 A_i ^c =0; \quad  A_i (t,{\bf x})=\Phi_i^{BPS}({\bf x}).
\ee
According to the theory of differential equations, the
general solution to a inhomogeneous equation can
be represented as the sum
\be
\label{inhom}
A_0^a={\cal Z}^a+{\tilde A}_0^a,
\ee
where ${\tilde A}_0^a$ is a particular solution to the
inhomogeneous equation being considered and ${\cal Z}^a $ is a solution
to the corresponding homogeneous equation
\be
\label{hom}
(D^2 (A))^{ab}{\cal Z}_b =0.
\ee
 The phase of
the  topological degeneration, $\hat\Phi _0 ({\bf x})$, is
such solution to the homogeneous  equation (\ref {hom}): apart from the factor $\dot N(t)$. Thus
\be
\label{Zm}
{\hat {\cal Z}}(t,{\bf x}) =
{\dot N}(t)\hat\Phi _{0}({\bf x}).
\ee
The solution to the homogeneous equation (\ref {hom}) describes
an "electric" monopole,
 \be
\label{el.m}
G^a_{i0}({\cal Z})= D_i ^{ab}(\Phi^{(0)BPS}){\cal Z}_b
= \dot N (t)D ^{ab}_i(\Phi ^{(0)BPS})\Phi_{0b},
\ee
which cannot be completely removed from the action functional of the theory being considered and from the Pontryagin
index by going over to Dirac's variables with the aid of
 gauge transformations (\ref {d.v.}). As was shown in \cite{David1}, the Pontryagin index
\be
\label{Pon}
\nu[A]=\frac {g^2}{16\pi ^2}\int \limits _{t_{in}} ^{t_{out}} dt\int \limits _V d^3 x G_{\mu \nu}^a
{\tilde G}^{\mu \nu}_a =N(t_{out})-N(t_{in}), \ee
depends only on the difference of the final and initial values of the topological degree of freedom. \par
In the lowest order of the perturbation theory in the
coupling constant
the action functional of the theory being considered contains, in addition to vacuum YM
BPS monopoles,
"electric" monopoles and describes the dynamics of the
new topological variable $N(t)$  in the form of a free
rotator. This induces the complete action for the (YM-Higgs) Bose condensate:
\be
\label{ca}
W_{\cal Z} [N,\Phi^{BPS}]= \int dt \int \limits_V d^3x  \frac {1}{2}\{ [G^b_{i0}({\cal Z}) ]^2-
[B^b_i(\Phi_k^{bBPS})]^2\}= \int dt\frac {1}{2}\{ I\dot N ^2-
\frac {4\pi}{g^2 \epsilon}\},
\ee
where
\be
\label{I}
I=\int \limits_ V ~d^3x (D_i^{ac}(\Phi_k^0)\Phi_{0}^c)^2 =
\frac {4\pi}{g^2} (2\pi )^2\epsilon
\ee
is the angular momentum of the rotator and $\epsilon= \sqrt{\lambda}/gm$ is the typical size of BPS monopoles. \par
The  Bose condensation action obtained in  such wise  specifies
the \bf Poincare invariant \rm  Hamiltonian of the BPS monopole vacuum in terms of the canonical momentum
$ P_N =\dot N I$:
\be \label{Hamilton} H= \frac {2\pi}{g^2\epsilon}[ P_N^2 (\frac {g^2}{8\pi^2})^2+1].
\ee
Upon  introducing  new Dirac's variables with the
aid of  transformations (\ref{d.v.}), the topological degeneracy of all  the fields reduces, for such Dirac's variables, to the degeneracy of only one topological variable $N(t)$ with
respect to
  shifts of this variable on integers:
($N \Longrightarrow N+n$, $n= \pm 1,\pm 2,...$). \par
The wave
function for the topological motion in the Minkowski
space has the form of the free-rotator wave function
\be
\label{Fmon}
\Psi _{mon}[N] =\exp \{iP_N N\},
\ee
with the momentum spectrum being determined from
the condition  $\Psi_{mon.} (N+1)=e^{i\theta}\Psi_{mon.}(N)$:
\be
\label{PsiN}
P_N ={\dot N} I= 2\pi k +\theta,
\ee
where $k$ is the number of a Brillouin zone and $\theta$ is the angle that specifies the spectrum of physical
values of the vacuum Hamiltonian (\ref{Hamilton}). This Hamiltonian has the zero eigenvalue ($H=0$) for the unphysical
momentum values $P_{ N}=\pm 8\pi i/g^2$, at which the
instanton wave function (\ref{tri}) coincides with the wave
function (\ref{Fmon}) for the monopole vacuum under the assumption that the topological degree of freedom is exclusively determined by a functional of local variables. \par
Thus the basic distinction between the monopole
vacuum and the instanton one is that in the first case  there arises an independent (pseudo)Goldstone mode
associated with the spontaneous breakdown of symmetry of physical states under the transformations of
the $ \pi_3 (SU(2))=\bf Z$ homotopies group. \par
The equation (\ref{PsiN}) and the
Bogomol'nyi equation (\ref{Bog}), which ensures
the potential character of the vacuum "magnetic" field $\bf B$, determine
the spectrum of the vacuum "electric" field ("electric" monopole):
\be
\label{el.m1}
G^a_{i0} =\dot N(t) ~(D_i (\Phi_k^0) \Phi_0)^a= P_N \frac {\alpha_s}{\pi^2\epsilon} B_i^a (\Phi _0)=\vert 2\pi k +\theta\vert \frac {\alpha_s}{\pi^2\epsilon}
B_i^a(\Phi _0).
\ee
The expression (\ref{el.m1}) is an analogue of the spectrum of the
electric field tension $$G_{10}=e(\frac {\theta}{2\pi}+k)$$
in two-dimensional electrodynamics \cite{Coleman,Ilieva2,Gogilidze1} characterized
by the same topology of degeneracy of initial data:
$$ \pi_1 (U(1))= \pi_3 (SU(2))=\bf Z.$$
The vanishing of the topological momentum does not
imply that the degeneracy of physical states disappears. Physical implications of such a degeneracy  will
be considered in the next section.
\section{Yang-Mils theory featuring   topological degeneracy
of
physical states.}
It is well known that the perturbation theory constructed for non-Abelian fields by analogy with
QCD \cite{Schwinger2,Fadd1} is infrared-unstable \cite{Mat,A.Vladimirov}. A conventional vacuum of the perturbation theory, $A=0$, is
not a stable state. As a rule, homogeneous (\cite{Mat})
or singular
fields, including instantons (\cite{Bel,Al.S.}),
the equation for
which involves delta-function-like singularities in the
Euclidean space $E_4$, are used for  describing nonzero vacuum fields.
If one explains physical effects by a homogeneous (or by
an instanton) vacuum,  it is also necessary to explain
the emergence of capacitors at the spatial infinity that
generate homogeneous fields (or the origin of sources of
delta-function-like singularities) \footnote{We recommend our readers the interpretation
of sources with the help of delta-function-like singularities given in
the monograph \cite{V.S.Vladimirov}: in \S 11.3.}. \par
Unlike, BPS
monopoles
provide a unique possibility for  introducing, in the
class of regular functions associated with topologically nontrivial gauge transformations, vacuum fields
in such a way that the equations of the Yang -Mills theory (it is (\ref {L}) in our case)  do not develop any additional
sources. \par
In order to introduce such a monopole vacuum,
we include the interaction of  gauge fields with a
Higgs field in a space of  the finite volume $V =\int d ^3 x$.
The scalar-field condensate forms a BPS
monopole
characterized by a
finite mass of the scalar field:
\be
\label{mass}
 \frac{1}{\epsilon}=\frac{gm}{\sqrt{\lambda}}=\frac{g^2<B^2>V}{4\pi}. \ee
If we go over to the infinite-volume limit $V \to \infty$ under
the condition that $<B^2> $ is finite, the scalar field acquires
an infinitely large mass and disappears from the spectrum of
physical excitations, while the regular solution represented by a
vacuum BPS monopole continuously  transforms into a Wu-Yang
monopole \cite{Wu}; this means that the equations of the theory do
not develop, at the origin of coordinates, a singularity inherent
in Wu-Yang monopoles \cite{Wu} and that the energy density does
not go to infinity. In this limit the finite energy density of the
appropriate BPS monopole has the form \be \label{Bint} \int d ^3 x
[B_i^a]^2 \equiv V <B^2>, \ee where $<B^2>$ is the quantity that
one has for the order parameter of the physical vacuum of the
gauge field upon the  removal of the scalar field. \par We recall
that  a Wu-Yang monopole \cite{Wu} is an exact solution to the
classical equations of the "pure" YM theory, (\ref{D}),
everywhere, with the exception of an infinite-small  neighbourhood
of the origin of coordinates. Just in this small region around the
origin of coordinates  Wu-Yang monopoles are regularized by the
scalar-field mass and this region disappears in the
infinite-volume limit $V \to \infty$ in (\ref{Bint}).\par It
should be noted here that, in quantum field theories, a transition
to the limit $V \to \infty$ is performed upon  calculating
physical observables, such as scattering cross sections and decay
probabilities, that are normalized per unit time and per unit
volume. Therefore, all the specific features of the above theory
involving  vacuum BPS monopoles,  which include the topological
degeneracy of initial data and "electric" monopoles,  survive at
any finite value of the volume. \par On the other hand, there are,
in the YM theory specified by the Lagrangian density (\ref{L}),
direct indications that the scale (gauge) symmetry of the vacuum
is broken by solutions belonging to the type of a Wu-Yang monopole
\cite{Wu}. In particular, the topological classification of the
classical solutions to the "pure" YM theory specifies the class of
solutions that possess the zero topological index ($n=0$), \be
\label{choose} X[A=\Phi^{(0)}]=0,\quad \frac {\delta X[A]}{\delta
A_i ^c} \vert _{A=\Phi^{(0)}} \neq 0, \ee and which have the form
\be \label{vec} {\hat\Phi}_i^{(0)} =-i \frac {\tau ^a}{2}\epsilon
_{iak}\frac {x^k}{r^2}f(r), \ee where there is only the one
unknown function, $f(r) $. The equation for this function can be
obtained by  substituting the expression (\ref{vec}) into the
classical equation (\ref{D}): \be \label{eq}
D^{ab}_k(\Phi_i^{(0)})G_{b}^{kj}(\Phi_i^{(0)})=0 \Longrightarrow
\frac {d^2f}{d r^2}+\frac {f (f^2-1)}{r^2}=0. \ee In the region
$r\neq 0$ there exist the following three solutions to this
equation: \be \label{3} f_1^{PT}=0,\quad f_1^{WY}=\pm 1. \ee The
first, trivial, solution $f_1^{PT}=0 $ corresponds to the ordinary
unstable perturbation theory involving the "asymptotic freedom "
\cite{Mat,A.Vladimirov}. Two nontrivial solutions \linebreak
$f_1^{WY}=\pm 1$   represent Wu-Yang monopoles, which, in the
model being considered, emerge from (vacuum) BPS monopoles in the
infinite-volume  limit without their singularities, with
(pseudo)Goldstone modes accompanying the breakdown of the scale
invariance. \par Thus the monopole vacuum characterized by a
topological degeneracy of all the physical states has the
following features distinguishing it from the topologically
degenerate instanton vacuum \cite{Bel,Hooft1,Al.S.}:  the
Minkowski space;  topological (pseudo)Goldstone modes associated
with the scale-symmetry breaking that generates the nonzero order
parameter $<B^2>\neq 0$ and a clear physical origin of the
scale-symmetry breaking, which is due to the condensate of the
scalar Higgs field and which survives upon  removing   the scalar
field from the excitation spectrum. \par Physical implications of
the theory being considered, which involves the YM
Min\-kow\-ski\-an monopole vacuum, are controlled by the
generating functional for the unitary perturbation theory in the
covariant Coulomb gauge. Reproducing the calculations performed in
the work \cite{Fadd1} (also see \cite{Nguen1}), one can obtain, as
the generating functional for such a perturbation theory, a
Feynman path integral in a reference frame with a specified time
axis, $l_\mu =(1,0,0,0)$: \be \label{Fe} Z^* [l,J^*] = \int\int
\prod \limits_ t ~d N(t) \prod  \limits_{c~=1}^{c~=3} [d^2
A^{*c}d^2 E^{*c}] \exp{\{iW^*[N, A^*,E^*]+i\int d^4x J^* \cdot
A^*\}}, \ee where $A^*$ are Dirac's variables (\ref{d.v.});  $E^*$
are their canonically conjugate momenta; $J^*$ are their sources;
$W^*[N, A^*,E^*]$ is the original action functional taken on the
manifold spanned by solutions to the constraint equation, \be
\label{Gauss2} \frac {\delta W}{\delta A_0}=0,\Longrightarrow
D_i^{cd}(A) G _{d0}^i =0, \ee for the non-Abelian electric-field
tension $ G _{0i}^d$ represented in the form of the sum of the
transverse momentum $E^*$ and  the longitudinal component: \be
\label{decomp} G ^d_{0i}= E^{*d}_i+ D_i^{db}(\Phi) \sigma _b \quad
( D_i^{cd}(\Phi^{N(WY)}) E_{d}^{*i}= 0 ). \ee If one assumes that,
in the perturbation theory, independent Dirac's variables
(\ref{d.v.}), $A_i^{*d}=\Phi_i^{dN(WY)}+\bar A_i^{*d}$, satisfy
the gauge condition \be \label{gc} D_i^{cd}(\Phi^{N(WY)}
A_{d}^{*i}= 0, \ee which were considered above, then the
constraint equation (\ref{Gauss2}) reduces to an equation for the
function $\sigma ^b$: \be \label{cur}
D_i^{cd}(A^*)D^i_{db}(\Phi^{N(WY)})\sigma ^b = j_0 ^c, \ee where
the quantity in the right-hand side is the current  of independent
non-Abelian variables, \be \label{cur1} j_0 ^a = g \epsilon ^{abc}
[ A_i^{*b}-\Phi_i^{b~N(WY)}]\tilde E ^{*i}_{c}, \ee belonging to
the excitations spectrum.   \par One can solve the equation
(\ref{cur}), which involves  transverse quantum excitations
$\tilde E_*$ over the zero mode (described above), by means of the
perturbation theory, employing a Green's function of the Coulomb
type. In the lowest order of the perturbation theory this Green's
function $G^{bc}({\bf x},{\bf y})$ in the field of an usual
Wu-Yang monopole $\Phi^{WY}$ is determined by the equation \be
\label{Gr.eq} [D^2(\Phi ^{WY})]^{ab}({\bf x})G_b^c ({\bf x},{\bf
y})= \delta^{ac}\delta^3({\bf x-y}). \ee A solution to this
equation specifies, in the Hamiltonian, an instantaneous
interaction of non-Abelian currents, \be \label{cint} -\frac
{1}{2} \int \sb{V_0} d^3 x d^3 y j_0 ^b ({\bf x}) G_{bc} ({\bf
x},{\bf y})j_0^c({\bf y}), \ee as an analogue  of the Coulomb
interaction of the currents in QED. A solution to the equation
(\ref{Gr.eq}) in the presence a Wu-Yang monopole, where \be
\label{Gr.eq.mon} [D^2(\Phi ^{WY})] ^{ab}({\bf x}) =
\delta^{ab}\Delta -\frac {n^a n^b+\delta^{ab}}{r^2}+2(\frac
{n^a}{r}\partial ^b- \frac {n^b}{r}\partial ^a); \ee
$n_a(x)=x_a/r$; $r=\vert {\bf x}\vert$, was obtained in the works
\cite{Pervush1,David2} by means of an expansion of $G^{ab}$ in
terms of a complete set of orthogonal vectors: \be \label{complete
set} G^{ab}({\bf x},{\bf y})= [n^a({\bf x}) n^b({\bf y})V_0(z)+
 \sum \limits_ {\alpha=1,2}  e^a_ \alpha (x)e^{b\alpha}(y)V_1(z)];\quad z=\vert {\bf x}- {\bf y}\vert,
\ee
where $V_{0,1}(z)$ are potentials. Substituting this expansion into the equation (\ref{Gr.eq}), one can derive an equation for
the potentials. The result is
\be
\label{Euler}
\frac {d^2}{dz} V_n+ \frac {2}{z}\frac {d}{dz}V_n- \frac {n}{z^2}V_n =0; \quad n=0,1.
\ee
Solving this equation, we obtain the potentials
\be
\label{V}
V_n (\vert {\bf x}-{\bf y} \vert)=
d_n\vert {\bf x}-{\bf y} \vert ^{l^n_1}+c_n\vert {\bf x}-{\bf y} \vert^{l^n_2}; \quad n=0,1,
\ee
where $d_n$ and $ c_n$ are constants, while $l^n_1$ and  $l^n_2$ are the
roots of the equation $(l^n)^2+l^n=n$, i.e.
\be
\label{roots}
l^n_1= -\frac {1+\sqrt{1+4n}}{2};~~~~~\quad l^n_2=\frac {-1+\sqrt{1+4n}}{2}.
\ee
At $n = 0$ we have $l^0_1= -(1+\sqrt{1})/{2}=-1$ and
$l^0_2=(-1+\sqrt{1})/{2}=0$,   so that there arises the Coulomb
potential
\be
\label{Cp}
V_0 (\vert {\bf x}- {\bf y} \vert) =
-1/ 4\pi \vert {\bf x}- {\bf y} \vert ^{-1} + c_0;
\ee
at $n = 1$, $l^1_1= -(1+\sqrt{5})/2\approx -1.618$ and $l^1_2=(-1+\sqrt{5})/2\approx 0.618 $, in  which  case one get the rising potential for the golden-section equation $(l^1)^2+l^1=1$:
\be
\label{V1}
V_1 (\vert {\bf x}-{\bf y} \vert)=
-d_1\vert {\bf x}-{\bf y} \vert ^{-1.618}+c_1\vert {\bf x}-{\bf y} \vert^{0.618}.
\ee
As was shown in the works \cite{Bogolubskaya,Yura2,Yura3},
the instantaneous interaction of colour currents through a rising potential
rearranges perturbation-theory series and leads to the
constituent mass of the gluon field in Feynman diagrams; this changes the asymptotic-freedom formula
at low momentum transferred, so that the coupling
constant $\alpha _{QCD}(q^2\sim 0)$
becomes finite. The rising
potentials of the instantaneous interactions of colour
currents \cite{Bogolubskaya,Yura2,Yura3} also lead to a spontaneous breakdown of the chiral invariance for quarks. \par
  Rising potentials do
not remove poles of Green's
functions in a perturbation theory for amplitudes of processes not involving colour degrees of freedom. Such a perturbation theory is formulated in terms of fields that are
characterized by zero topological quantum numbers
and  that can be called partons:
\be
\label{part}
\hat A^* (N\vert A^{(0)})= U_N[\hat A^{(0)}+\partial ]U^{-1}_N.
\ee
By virtue of the gauge invariance, the phase factors
of the topological degeneracy,
$$ U_N =\exp \{N(t)\hat \Phi_0 ({\bf x})\},$$
disappear. However, these factors survive at the
sources of physical fields in the generating functional (\ref{Fe}).
A theory featuring a topological degeneracy of initial data, where the sources of physical fields
involve the Gribov factors
$${\rm tr} [\hat J^i v^{(n)}\bar{\hat A}_i^{(0)}v^{(n)-1}]$$
differs from a theory that is free from any degeneracy
and  that involves the sources ${\rm tr} [\hat J^i\bar{\hat A}^{(0)}_i]$.   In a theory
featuring a degeneracy of initial data it is necessary
to average amplitudes over degeneracy parameters.
Such averaging may lead to the disappearance of a
number of physical states. \par
In \cite{Pervush4,Nguen2}
it was shown that amplitudes for the
production of physical colour particles may vanish
because of the destructive interference between the
phase factors of  the topological degeneracy. In this case
the probability-conservation law for the $S$-matrix elements \linebreak $ <i\vert S=I+i T\vert j>$ in the form
$$\sum \limits_f <i\vert  T\vert f>< f\vert  T^*\vert j> =
2 ~ {\rm Im} <i\vert T \vert j>$$
is exclusively saturated by the production of colour-singlet states (hadrons) $f=h$. By virtue of the
probability-conservation law, the sum over all the hadronic channels becomes equal to the doubled imaginary
part of the colour-singlet amplitude $2 ~ {\rm Im} <i\vert T \vert j>$.  \par
In
turn, the dependence on the factors of the topological
degeneracy completely disappears  in the colour-singlet
amplitude. Owing to the gauge invariance, the Hamiltonian of the theory, $H[A^{(n)}]=H[A^{(0)}]$,
depends
only on the fields of the zero topological sector, $A^{(0)}$, which play the role of Feynman 's partons. In
the high-energy parton region, where the imaginary
part of the colour-singlet amplitude, ${\rm Im} <i\vert T \vert j>$, can
be calculated on the basis of the perturbation theory,
the quark-hadron duality, which is used to directly measure
 parton quantum numbers coinciding with the
quantum numbers of physical colour particles, arises
from the probability-conservation law.
\section{Estimating   quantity $<B^2>$ within QCD.}
Let us estimate the quantity $<B^2>$ within QCD. \par
In the monopole vacuum of QCD the antisymmetric Gell-Mann matrices $\lambda_ 2,\lambda_5,\lambda_7$ play the role of the
matrices $\tau_1,\tau_2,\tau _3 $.  \par
The instantaneous interaction
of colour quark currents through a rising potential
leads to the spectrum of mesons:  in particular, to
the pseudoscalar $\eta _0$ meson. Its anomalous interaction
with gluons is described in terms of the Veneziano
effective action \cite{Veneziano}
\be
\label{inter}
W_{eff}=\int dt \left\{\frac 1 2 \left({\dot\eta_0}^2-M_0^2{\eta_0}^2\right)
 V +
 C_0\eta_0 \dot X[A^{(N)}] \right\},
\ee
in the rest frame of this meson. Here $V$ is the spatial volume; $C_{0}=(N_f/F_\pi)\sqrt{2}/ \pi $ is the coupling
constant for the anomalous interaction of the meson with
the topological functional $X[A^{(N)}]= X[A]+N$;
$F_\pi$ is the weak-pion-decay constant and $N_f=3$ is the
number of flavours. \par
The calculation of a similar action
functional for QCD and for QED$_{(3+1)}$, where the
topological functional describes the decay of a parapositronium into two photons, is represented in the work \cite{Pervush1}. \par
 In all probability, the expression (\ref{inter})  for the effective
anomalous interaction of a pseudoscalar state with
gauge fields is common for all the gauge theories. \par
For
electrodynamics in the two-dimensional space-time one
can obtain the same effective action \cite{Ilieva2,Gogilidze1}, where it
leads to the mass of the Schwinger bound state. \par
In QCD$_{(3+1)}$   the extra mass of a bound pseudoscalar state,
$$\Delta m_\eta^2 =C_0^2 /I_{QCD}V,$$
can be determined, upon adding the action functional
that is specified by  the formulas (\ref{ca}), (\ref{I}) and that controls the topological dynamics of the zero mode,
\be
\label{el.f}
W_{ QCD} =\frac{1}{2}\int dt  \int \limits_ V d^3 x G^2_{0i}=
 \int dt\frac{\dot N^2 I_{ QCD}}{2} \quad
(I_{ QCD}=\left(\frac{2\pi}{\alpha_{QCD}}\right)^2\frac{1}{V<B^2>}),
\ee
to the anomalous Veneziano action, by  diagonalizing
the total Lagrangian:
\be
\label{La}
L=[\frac{\dot N^2I_{QCD}}{2}+C_0\eta_0 \dot N] =
[\frac{(\dot N+C_0\eta_0/I_{QCD})^2I_{ QCD}}{2}- \frac{C_0^2}{2I_{ QCD}} \eta_0^2].
\ee
In QED$_{(1+1)}$
the analogous formula describes the
mass of the Schwinger state \cite{Ilieva2,Gogilidze1}, whereas in  QCD$_{(3+1)}$ we obtain the extra mass of the  $\eta_0$ meson:
\be
\label{leff}
 L_{\rm eff}= \frac{1}{2}[{\dot\eta_0}^2-\eta_0^2(t)(m_0^2
 +\triangle m_{\eta}^2)]V,
\ee
\be
\label{triangle}
\triangle {m_\eta}^2=\frac{C_{\eta}^2}{I_{ QCD}V}
 = \frac{N_f^2}{F_{\pi}^2}\frac{\alpha_{QCD}^2<B^2>}{2\pi^3}.
\ee
This result makes it possible to assess the  vacuum expectation value
of the chromo-magnetic field in QCD$_{(3+1)}$:
$$<B^2>\alpha_{QCD}^2=\frac{2\pi^3 F_{\pi}^2\triangle
m_\eta^2}{N_f^2}=0.06~ {\rm GeV}^4,$$
by
using the estimate $\alpha_{QCD}(q^2 \sim 0)\sim 0.24$  \cite{Bogolubskaya,David3}. \par
Upon the calculation, we can remove infrared regularization by going over to the limit
$V\to \infty$.
\section*{ Conclusion.}
\renewcommand{\theequation}{C.\arabic{equation}} \setcounter{equation}{0}
The monopole-vacuum model considered here
demonstrates that the quantization of a non-Abelian
theory featuring a topological degeneracy of the initial
data for all the physical states in a specific reference
frame describes a destructive interference of degeneracy phase factors, which leads to the quark-hadron
duality; a (pseudo)Goldstone mode that is associated with
a spontaneous breakdown of the initial gauge symmetry and
which leads to an extra mass of the $\eta_0$
meson;
a rising potential that controls the instantaneous interaction of
currents and which is thought to be responsible
for a spontaneous breakdown of the chiral symmetry.
These hidden features of non-Abelian fields manifest
themselves upon switching on and off  the gauge field
interaction with a Higgs field, which acquires an
infinitely large mass in the  infinite-volume limit.
There arises the question of the extent to which such
a fantastic possibility may be realized in nature. \par
The generating functional found in the form of a
Feynman path integral with respect to Dirac's variables can be recast  \cite{Fadd1}
into the form of a FP integral \cite{FP1} by means of the change of variables
$${\hat A}^*_i(N\vert A)= (U_n U^D [A]) [{\hat A}_i +\partial _i](U_n U^D[A] )^{-1},$$
where
\cite{Pervush1}:
\be
\label{UD}
U^D[A]= \exp \{\frac {1}{D^2(\Phi^{WY})} D_k (\Phi^{WY}) {\hat A}^k\}
\ee
is  Dirac's "dressing" of non-Abelian fields. Upon this
change of variables, the Feynman path integral is reduced to
the FP integral
\newpage
\bea
\label{ymfpi}
 Z[l,J^{*}]&=&\int \prod \limits_{t} dN(t)  \int\int \prod\limits_{c=1 }^{c=3 }
 [d^4A^{c} ]\delta(f(A)) ~ {\rm Det}~ M_{FP} \nonumber\\
 &&\times\exp\left\{i W [A]+i\int
 d^4x J^{*} \cdot A^{*}(N\vert A)\right\}
 \eea
in an arbitrary gauge of the physical variables, $f(A)=0$; here the FP determinant ${\rm Det}~ M_{FP}$ is determined in terms of the linear response of this gauge to a
gauge transformation, $f (e^\Omega (A+\partial)e^{-\Omega} =M_{FP}\Omega +
O(\Omega^2)$, while $W$ is the original action functional in
the theory being considered. At the same time, there
remains the Dirac gauge of the sources in (\ref {UD}).
\par
As
a relic of the fundamental quantization, the Dirac
phase factors in the integral in (\ref {ymfpi})  "remember " the
entire body of information about the reference frame;
monopoles; the rising potential of the instantaneous
interaction and other initial data, including their
topological degeneracy and confinement. \par
As was predicted by Schwinger \cite{Schwinger2}, all these effects disappear,
leaving no trace, if these Dirac factors are removed
by means of the  replacement $A^{*}(N\vert A)\Longrightarrow A$ \cite{Fadd1},
which is made with the only purpose of removing the
dependence of the path integral on a reference frame and
initial data. On getting rid of this dependence,
we obtain, instead of hadronization and confinement
in the Dirac's quantization scheme, only the amplitudes for
the scattering of free partons in the "relativistic "
FP integral, which do not exist as physical observables in the initial-data-dependent Dirac's
scheme of quantization. \par
The same metamorphosis occurs in QED as well:
going over from the Dirac gauge of sources to the
Lorentz gauge in order to remove the dependence on a
reference frame and initial data, we replace the perturbation theory emerging upon the fundamental quantization
and featuring two singularities in the photon propagators (a single-time singularity and that at the light
cone) by the perturbation theory in the Lorentz gauge; the
latter involves only one singularity in the propagators
(that at the light cone), but, in principle, cannot
describe single-time Coulomb atoms, containing only
Wick-Cutkosky bound states whose spectrum in not
observed in nature \cite{Kummer}.
\section*{Acknowledges.}
We grateful to B. M. Barbashov, D. Blaschke,
E. A. Kuraev and V. B. Priezzhev for discussion.
V. N. Pervushin is indebted to W. Kummer for providing information about Wick-Cutkosky bound states
and to A. S. Schwartz for a critical comment.

\end{document}